\documentclass[12pt,preprint]{aastex} \usepackage{apjfonts}
\usepackage{amsmath}
\usepackage{epsfig}
\usepackage{graphicx,epsfig}
\usepackage{natbib}
\newcommand\lsim{\mathrel{\rlap{\lower4pt\hbox{\hskip1pt$\sim$}}
        \raise1pt\hbox{$<$}}}
\newcommand\gsim{\mathrel{\rlap{\lower4pt\hbox{\hskip1pt$\sim$}}
        \raise1pt\hbox{$>$}}}

\newcommand{\dd}{\partial}

\newcommand{\cm}{\;\mathrm{cm}}

\newcommand{\s}{\mathrm{s}}
\newcommand{\erg}{\;\mathrm{erg}}

\newcommand{\yr}{\;\mathrm{yr}}

\newcommand{\Msol}{M_{\odot}}

    \def\dd{\partial}

    \def\beq{\begin{equation} }
    \def\eeq{\end{equation} }
    \def\spose#1{\hbox to 0pt{#1\hss}}
    \def\ltsim{\mathrel{\spose{\lower.5ex\hbox{$\mathchar"218$}}
     \raise.4ex\hbox{$\mathchar"13C$}}}

\def\spose#1{\hbox to 0pt{#1\hss}}
\def\lta{\mathrel{\spose{\lower 3pt\hbox{$\mathchar"218$}}
        \raise 2.0pt\hbox{$\mathchar"13C$}}}
\def\gta{\mathrel{\spose{\lower 3pt\hbox{$\mathchar"218$}}
        \raise 2.0pt\hbox{$\mathchar"13E$}}}
\defcitealias{MP05}{MP05}
\defcitealias{TM10}{TM10}

\shorttitle{Witnessing the Birth of a Quasar}
\shortauthors{Tanaka et al.}

\begin{document}

\title{Witnessing the Birth of a Quasar}

\author{Takamitsu Tanaka, Zolt\'an Haiman \& Kristen Menou\footnote{
Kavli Institute for Theoretical Physics, UCSB, Santa Barbara, CA 93106}}

\affil{Department of Astronomy, Columbia University, 550 West 120th
Street, New York, NY 10027}

\begin{abstract}
  The coalescence of a supermassive black hole binary (SMBHB) is
  thought to be accompanied by an electromagnetic (EM) afterglow,
  produced by the viscous infall of the surrounding circumbinary gas
  disk after the merger.  It has been proposed that once the merger
  has been detected in gravitational waves (GWs) by the {\it Laser
  Interferometer Space Antennae} ({\it LISA}), follow-up EM
  observations can search for this afterglow and thus help identify
  the EM counterpart of the {\it LISA} source.  Here we study whether
  the afterglows may be sufficiently bright and numerous to be
  detectable in EM surveys alone.  The viscous afterglow is
  characterized by an initially rapid increase in both the bolometric
  luminosity and in the spectral hardness of the source.  For binaries
  with a total mass of $10^{5}-10^{8}\Msol$, this phase can last for
  years to decades, and if quasar activity is triggered by the same
  major galaxy mergers that produce SMBHBs, then it could be
  interpreted as the birth of a quasar.  Using an idealized model for
  the post-merger viscous spreading of the circumbinary disk and the
  resulting light curve, and using the observed luminosity function of
  quasars as a proxy for the SMBHB merger rate, we delineate the
  survey requirements for identifying such birthing quasars.  If
  circumbinary disks have a high disk surface density and viscosity,
  an all-sky soft X-ray survey with a sensitivity of $F_{\rm X}\ltsim
  3\times 10^{-14}~{\rm erg~s^{-1}~cm^{-2}}$ which maps the full sky
  at least once per several months, could identify a few dozen birthing
  quasars with a brightening rate $d\ln F_{\rm X}/dt > 10\% \yr^{-1}$
  maintained for at least several years.  If $>1\%$ of the X-ray
  emission is reprocessed into optical frequencies, several dozen
  birthing quasars could also be identified in optical transient
  surveys, such as the {\it Large Synoptic Survey Telescope}.
  Distinguishing a birthing quasar from other variable sources may be
  facilitated by the monotonic hardening of its spectrum, but will
  likely remain challenging.  This reinforces the notion that
  observational strategies based on joint EM-plus-GW measurements
  offer the best prospects for the successful identification of the EM
  signatures of SMBHB mergers.
\end{abstract}

\section{Introduction}

Observational evidence robustly indicates that all or nearly all
galaxies harbor a supermassive black hole in their nucleus
\citep[SMBH; e.g.,][]{Maggor+98}.  Since cosmological structure
formation models predict a hierarchy of galaxy mergers, if nuclear
SMBHs were indeed common at earlier times, then these mergers should
result in the formation of SMBH binaries
\citep[SMBHBs;][]{Begelman+80}, and these binaries should then be
common throughout cosmic time \citep{Haehnelt94, Menou+01, VHM03,
WL03, Sesana+07, Lippai+09, TH09}.

It has also long been known, both observationally (e.g.,
\citealt{Sanders+88}) and theoretically (e.g.,
\citealt{BarnesHernquist+91}) that galaxy mergers can drive gas to the
nucleus of the merger remnant, which could facilitate the merger of
the nuclear SMBHs on one hand, while also providing fuel for quasar
activity on the other.  Mergers are therefore generically also
believed to trigger quasar activity; the rate of major galaxy mergers
can indeed provide an explanation for the observed evolution of the
quasar population as a whole (\citealt{Carlberg+90}; for more recent
work, see, e.g., \citealt{Hopkins+07a} and \citealt{WyitheLoeb+09} and
references therein).

Despite their expected ubiquity, observational evidence for SMBHBs is
scarce, and the precise timing of any quasar activity, and when it
occurs relative to the merger of the nuclear SMBH binary, remains
unclear \citep{Kocsis+06}. A handful of pairs of active SMBHs in the
same galaxy have been resolved directly, at $\sim$kpc separation in
X--ray \citep{Komossa+03} and optical \citep{Comerford+09} images, and
at $\sim$10pc separation in the radio \citep{Rodriguez+06}, confirming
that gas is present around the SMBH binary, and that quasar activity
can, at least in some systems, commence prior to their coalescence.
However, there has been at least one suggestion that luminous activity
can be occurring later, at the time of the merger, as well --
momentarily interrupted by the coalescence of the SMBHs and
reactivated after-wards \citep{Liu+03}.  While there are many more
observed SMBHB candidates with small separations
\citep[e.g.,][]{Roos+93, Schoen+00, Merritt+02, Sudou+03, Liu04,
Boroson+09}, the evidence for these tighter binaries is indirect, and
each candidate system has alternative explanations.  The expectation
is that at large separations, the binaries rapidly lose orbital
angular momentum through dynamical friction with background stars and
through tidal--viscous interaction with the surrounding gas
\citep{IPP99, AN02, Escala+05b, MM05, Dotti+07, SHM07, Cuadra+09,
Callegari+09, Colpi+09, HKM09, Chang+09}.  Once sufficiently compact,
gravitational wave (GW) emission rapidly shrinks the orbit,
culminating in a merger.  How long this process lasts, and at what
stage(s) the SMBHs light up as luminous quasars, is, however, also
poorly understood theoretically.

Apart from the cosmological context, interest in EM signatures of SMBH
mergers surged recently \citep[e.g.,][]{MP05, BP07, Lippai+08, SK08,
SB08, O'Neill+09, Chang+09, Megevand+09, Corrales+10, Rossi+10,
Anderson+10, Krolik10, TM10, Shapiro10}, driven by (i) the prospect
that the {\it Laser Interferometer Space
Antennae}\footnote{http://lisa.nasa.gov/} ({\it LISA}) will detect the
mergers in GWs and provide a tractable list of (perhaps as low as a
few hundred; e.g., \citealt{Kocsis+08}) EM candidates for SMBHBs and
(ii) the breakthrough in numerical general relativistic calculations
of BH mergers (e.g., \citealt{Pretorius05, Campan+06, Baker+06}),
which led to robust predictions of significant mass--loss and recoil
that can significantly perturb the ambient gas.  A simultaneous
observation of the merger in gravitational and EM waves would enable
new scientific investigations in cosmology and BH accretion physics
\citep{Cutler98, HH05, Kocsis+06, Kocsis+07, LH08, Phinney09,
Bloom+09}.

In this paper, we focus on one particular signature of SMBHB
coalescence, which we will hereafter refer to as the ``viscous
afterglow''.  The physics of this model was discussed by
\citet{Liu+03} in the context of the interruption of jets in
double-double radio galaxies, and later by \citeauthor{MP05} (2005;
hereafter MP05) in the context of EM counterparts of {\it LISA}
sources.  Prior to merger, the SMBHB torques open and maintain a
cavity in the center of a thin circumbinary gas disk
\citep{ArtymowiczLubow94}.  When the binary becomes sufficiently
compact, GW emission causes the binary orbit to shrink faster than the
gas just outside the cavity can viscously respond.  The merger takes
place inside the cavity, which is subsequently filled as the disk
viscously spreads inward.  Because the refilling inner disk produces
higher-energy photons than the outer regions, the disk is predicted to
transition from an X-ray-dim state to an X-ray-bright one, with its
bolometric luminosity increasing by a factor of $\sim 10$ during this
time.  This transition is expected to take place on humanly tractable
timescales, with the cavity filling in $\sim 10 (1+z)
(M/10^{6}\Msol)^{1.3} \yr$, where $M$ is the total mass of the binary.
A study of an optically selected sample by \cite{Gibson+08} found
X-ray-dim AGN to be rare ($\ltsim 2\%$ at $z\sim2$), suggesting that
it would be tractable to catalog and monitor such systems for possible
observational signatures of a merger afterglow.

In the observational scenario originally proposed by MP05, {\it LISA}
would detect the GWs from the merger and determine its approximate
location in the sky to within $\sim 0.1$ deg, triggering a follow--up
search to identify the EM counterpart and host galaxy.  A natural
question to ask, however -- and the subject of the present paper -- is
{\em whether the viscous afterglows may be sufficiently bright and
numerous to be detectable in EM surveys alone, even before {\it LISA}
is launched}.  The identification of mergers by their EM signatures
alone could, in fact, be valuable for several reasons.  First, {\it
LISA} will be sensitive to GWs from relatively low--mass SMBHBs, with
total masses of $\sim (10^{4}-10^{7})/(1+z)\Msol$.  EM studies could,
in principle, detect coalescing SMBHBs outside this mass range, and
therefore complement the {\it LISA} binary population.  Second, while
many models for the cosmological evolution of SMBHs predict that {\it
LISA} will detect dozens or hundreds of mergers (if ``seed'' black
holes are abundant and merge often; e.g., \citealt{Sesana+07}), there
are some SMBH assembly scenarios that may result in far fewer {\it
LISA} events (i.e. if seeds are rare and grow primarily through rapid
accretion or are very massive already at formation;
\citealt{TH09,Lippai+09}).  It is therefore plausible that EM surveys
could deliver a larger SMBH binary sample than available from GWs.
Third, several transient EM surveys are already under way, or are
planned to be completed before the expected launch date of the {\it
LISA} mission around 2020.

If luminous quasar activity is triggered by major mergers of galaxies,
as argued above, then the viscous afterglow could plausibly be
interpreted as the signature of the birth of a quasar.  In this paper,
we estimate the number of identifiable afterglow sources,
i.e. birthing quasars, in the sky, by (i) adopting an idealized
time-dependent model \citep[][hereafter \citetalias{TM10}]{TM10} of
the evolution of the disk structure, to calculate photometric light
curve and variability of the afterglow, and (ii) by using the observed
luminosity function of quasars as a proxy for the SMBHB merger
rate. Our two main goals are:

\begin{enumerate}

\item To assess whether there is any hope of detecting and identifying
the viscous afterglows with conventional EM telescopes alone.

\item To see how the identifiability of the afterglows depends on
theoretical parameters and to delineate the ideal survey attributes
(wavelength, angular coverage and depth).  We compare the derived
attributes to those similar to planned large surveys of the transient
sky: a soft-X-ray survey with specs similar to those that were
proposed recently, unsuccessfully, for the the {\it Lobster-Eye
Wide-Field X-ray Telescope}
\footnote{http://www.star.le.ac.uk/lwft/} ({\it LWFT}) mission;
and the {\it Large Synoptic Survey Telescope}\footnote{http://www.lsst.org/lsst} ({\it LSST}) in the optical.
\end{enumerate}

We find that the detectability of the afterglow is sensitive to the
properties of the circumbinary disk, in particular to the ratio of the
viscous stress to the gas pressure, and to the surface density of the
disk.  We conclude that purely EM identification of the afterglows by
the planned surveys are unlikely, unless the surface density and the
viscosity in the circumbinary disk are at the high end of the expected
range.  In this latter, optimistic scenario, several dozen birthing
quasars could be identified in a soft X-ray transient survey.  We also
find that if $\gsim 1\%$ of the X-ray radiation emitted in the central
regions is reprocessed into the optical frequencies by dust
surrounding the source, or by warps or geometric splaying in the disk
itself \citepalias{TM10}, several dozen afterglows could be detected
in an optical transient survey, such as {\it LSST}.

This paper is organized as follows.  In \S~2, we summarize the
viscous afterglow model, and describe our methods for estimating the
identifiable population of AGN harboring a recently merged SMBHB.  In
\S~3, we present estimates for the number of identifiable afterglow
sources in the sky.  We summarize our results and offer our
conclusions in \S~4.

\section{A Simple Model for the Afterglow Population}

In this section, we describe the model and underlying assumptions used
to estimate the number of observable afterglow sources in the sky.  We
use the idealized Newtonian viscous evolution model described by
\citetalias{TM10} to calculate the light--curve and spectral evolution
of the source, and to obtain light curves in fixed frequency bands.
Then, following \cite{HKM09} we assume that every SMBHB merger
ultimately leads to a bright quasar episode.  Using the bolometric
quasar luminosity function of \cite{Hopkins+07b} as a proxy for the
underlying SMBH mass function, we estimate the number of afterglow
sources in the sky that exhibit identifiable photometric brightening.
\footnote{ As discussed by \cite{HKM09}, an alternative method to
  construct the luminosity function is to calculate the SMBH merger
  rate from the merger rate of dark matter halos, and apply the
  post-merger light curve $L(t,M,q)$ to each merger event.  That
  method requires a third ingredient, namely a way to prescribe the
  SMBH masses from the halo masses.  Given the approximate nature of
  our calculation, we choose the simpler method described in the text.
}

Throughout this paper, $M=M_{1}+M_{2}$, $q=M_{1}/M_{2}$ and $a$
denote the binary mass, mass ratio and semi-major axis, respectively.
The physical constants $G$, $c$, and $k_{\rm B}$ have their usual
meanings.

\subsection{Modeling Afterglow Light Curves}
\label{subsec:curve}

Below, we briefly recapitulate the main features of the viscous
afterglow model, which was advanced by \citetalias{MP05} and
elaborated upon by \citetalias{TM10} (see also \citealt{Shapiro10}).
We refer the reader to those earlier works for further details; a
derivation of the disk evolutionary equations, in particular, can be
found in Appendix~A of \citetalias{TM10}.

A SMBHB in a circular orbit of semi-major axis $a$ will open an
annular gap in a thin circumbinary disk at a radial distance $R_{\rm
wall}\sim 2a$ from the binary's center of mass \citep{Artymo+91}.  The
binary will shrink gradually by depositing orbital angular momentum in
the disk, maintaining a nearly self-similar geometry with $R_{\rm
wall}/(2a)\sim 1$.  The kinematic viscosity $\nu_{\rm gas}$ of the gas
comprising the disk is a weak function of radius $R$, and thus the
viscous time $t_{\rm visc}(R)=(2/3)R^{2}/\nu_{\rm gas}$ roughly scales
just outside the cavity as $t_{\rm visc}(R_{\rm wall})\propto a^{2}$.
Because GW emission shrinks the orbit on a timescale $t_{\rm GW}\equiv
(d\ln a/dt)^{-1}\propto a^{4}$, there exists a value of $a$ inside
which the binary orbit begins to shrink faster than the circumbinary
gas can viscously follow.  This critical binary separation is of the
order $a\sim 100 GM/c^{2}$; past this point, the evolution of the
binary and the disk are decoupled.

After decoupling, the evolution of the disk surface density $\Sigma$
can be described by the standard equation for a thin, Keplerian
accretion disk \citep[e.g., ][]{FKR02}:
\beq
\frac{\dd}{\dd t}\Sigma (R,t)=\frac{3}{R}\frac{\dd}{\dd R}\left[
R^{1/2}\frac{\dd}{\dd R}\left(\nu\Sigma R^{1/2}\right)\right].
\eeq
In the special case when the dependence of the gas viscosity on radius
can be approximated as a power--law, $\nu\propto R^{n}$, the surface
density evolution $\Sigma (R,t)$ can be solved semi-analytically with
a Green's function formalism (\citealt{LP74}, \citetalias{TM10}),
starting from an arbitrary initial distribution $\Sigma(R, t=0)$.

The subsequent evolution of the circumbinary disk is sensitive to
several properties of the binary-plus-disk system.  For a given total
binary mass $M\equiv M_{6}\times10^{6}\Msol$, the relevant observable
quantities can be expressed in terms the following eight system
parameters.

\begin{itemize}

\item The parameter $\zeta \equiv 4q (1+q)^{-2}\le 1$ is the symmetric
  mass ratio of the binary, scaled to unity for equal-mass binaries.

\item The ratio of the diameters of the circumbinary cavity and the
  binary separation at decoupling is specified by the parameter
  $\lambda=R_{\rm wall}/(2a)\sim 1$.

\item The gas viscosity is parametrized via the ratio of the viscous
  stress to the gas pressure, $\alpha=\nu_{\rm gas}\Omega k_{\rm
  B}T/P_{\rm gas}$ \citep{SS73}.  The choice to scale the viscosity
  with the gas pressure and not the total pressure is motivated by
  calculations that suggest that radiation pressure-dominated thin
  disks may be thermally unstable \citep{SS76, Pringle76}.\footnote{
  Even if thermally stable \citep{HBK09}, such disks may still be
  viscously unstable \citep{LE74, Piran78, HBK09}.}

\item The parameter $\beta$ gives the ratio at decoupling of the
  timescale $t_{\rm GW}$ on which the binary shrinks due to GW
  emission to the viscous time $t_{\rm visc}$ of the gas at the cavity
  wall.  \citetalias{MP05} prescribed $\beta=0.1$, based on the fact
  that the gas near the cavity wall has a very steep density gradient
  and will move on a characteristic time--scale $\sim 0.1t_{\rm visc}$
  after the binary torques vanish.  However, the calculations by
  \citetalias{TM10} suggest that the lower value of $\beta\sim 0.05$
  is more appropriate, as it gives a closer approximation to the time
  elapsing between the merger and when the binary torques cease to
  influence the gas.

\item The disk porosity parameter, $\theta$, prescribes how the
  optical thickness $\tau$ between the disk midplane and the height
  where the emitted photons are thermalized relates to the surface
  density: $\tau=\theta\Sigma\sigma_{\rm es}$, where $\sigma_{\rm es}$
  is the electron-scattering cross section.

\item The viscous evolution model assumes a radial power--law for the
  kinematic viscosity, $\nu_{\rm gas}\propto R^{n}$.  Prior to merger,
  viscosity in a circumbinary decretion disk may be expected to be a
  weak function of radius, with $n\ltsim 0.1$; after the merger,
  $n\sim 0.4$ may be expected in the accretion region of interest
  \citepalias{TM10}.

\item Another parameter of interest is the radial power-law index
  $m\equiv \dd(\ln \nu\Sigma)/\dd \ln R$ of the disk mass profile at
  decoupling.  The standard steady-state solution for a thin accretion
  disk around a single central object satisfies $m=0$, with mass flow
  $\dot{M}=3\pi \nu_{\rm gas}\Sigma$ constant with radius.  However,
  at the cavity wall the binary torques {\it decrete} the gas outward;
  in this regime, the disk would more likely satisfy $m=-1/2$, with
  torque density $\nu_{\rm gas}\Sigma R^{1/2}~=~{\rm constant}$
  \citep[e.g., ][]{Pringle91}.  We prescribe initial surface density
  profiles satisfying
\[
\Sigma(R\ltsim R_{\rm wall})~\ll~ \Sigma(R\gtrsim R_{\rm wall})~\propto ~ R^{-n+m},
\]
with a steep exponential drop-off in $\Sigma$ near $R\approx R_{\rm
  wall}$ (\citealt{MM08}; \citetalias{TM10}).

\item The value of the surface density $\Sigma_{\rm wall}$ at the
  cavity wall at decoupling depends on how much gas has piled up due
  to accretion of outer gas and decretion of the gas just inside.  We
  parametrize this value as $\Sigma_{\rm wall}=S\dot{M}_{\rm
  Edd}/(3\pi\nu)$, where $\dot{M}_{\rm Edd}$ is the accretion rate
  that would produce a luminosity corresponding to the Eddington limit
  for the binary mass $M$, assuming a radiative efficiency of
  $\eta=10\%$.  The parameter $S$ can be thought of as the product of
  two quantities: the mass supply rate $\dot{M}$ to the disk in
  Eddington units, and the enhancement of the disk surface density
  just outside the cavity due to mass accumulation.  The second factor
  may be expected to exceed unity.

\end{itemize}

In addition to the above parameters, the cosmological redshift $z$ of
the source will also obviously affect the observed spectra,
luminosity, and brightening rates of the sources.

For a given set of parameters, we use the corresponding surface
density evolution $\Sigma (R,t)$ to calculate the disk temperature
profile at depths where the emitted photons are thermalized, which in
turn is used to obtain the monochromatic luminosity in the source rest
frame:
\beq
L_{\nu}(t)\sim 2\times \int_{R_{\rm ISCO}}^{10 R_{\rm wall}} \pi\frac{2\epsilon_{\nu}}{1+\epsilon_{\nu}}B_{\nu}~2\pi R~dR.
\label{eq:Lnu}
\eeq
The leading factor of two on the right hand side of equation
\eqref{eq:Lnu} accounts for the fact that the disk radiates from two
faces; $B_{\nu}$ is the Planck function; $\epsilon_{\nu}\equiv
\kappa_{\rm abs,\nu}/(\kappa_{\rm abs,\nu}+\kappa_{\rm es})$ is the
ratio of the absorption to the total opacity; and the fraction
$2\epsilon_{\nu}/(1+\epsilon_{\nu})$ is the deviation of the flux from
blackbody \citep[e.g.,][]{RL86, Blaes04}.  The effective temperature
in the inner region of the disk is higher than a blackbody disk with
the same surface density distribution, and thus it produces a harder
spectrum.  The lower limit of integration, $R_{\rm ISCO}$, is the
radius of the innermost stable circular orbit, for which we adopt the
value $3GM/c^{2}$, consistent with the assumption that the binary
remnant has moderate spin, $0.6\ltsim a_{\rm spin}\ltsim 0.9$.  The
choice for the upper limit of integration is somewhat arbitrary, and
does not significantly affect $L_{\nu}(t)$ as long as it is
sufficiently large; at larger radii the flux is significantly lower,
and evolves on much longer timescales than the inner region originally
occupied by the cavity.

The overall disk evolution timescale is roughly the viscous time at
the cavity wall, evaluated at decoupling:
\beq
t_{\rm visc}\sim 120\yr \times
M_{6}^{1.32}\zeta^{0.70}\lambda^{2.8}\alpha_{-1}^{-1.36}\beta_{-1}^{0.70}S^{-0.98}\theta_{0.2}^{-0.34}.
\label{eq:tvisc}
\eeq
Although the disk continues to brighten and spectrally harden for
$\sim t_{\rm visc}$ after the merger, the most dramatic evolution
takes place in the first $\sim \beta t_{\rm visc}$, which corresponds
to the faster viscous spreading of the sharp density edge at $ R \lsim
R_{\rm wall}$.

The time--dependent spectrum of the viscously spreading disk can be
divided into two parts: a nearly static low--frequency component
produced predominantly by gas outside $R_{\rm wall}$; and a rapidly
evolving high--frequency component produced predominantly by the gas
flowing into the central cavity.  The characteristic frequency that
marks the boundary between the static and variable parts of the
spectrum can be approximated in the source rest frame
\citepalias{TM10} as
\beq 
h\nu_{\rm var}\sim 30~{\rm eV} \times
M_{6}^{-0.32}\zeta^{-0.45}\lambda^{2.1}\alpha_{-1}^{0.36}\beta_{-1}^{-0.45}S^{0.73}\theta_{0.2}^{0.09},
\label{eq:hnuvar}
\eeq
or, in terms of the wavelength,
\beq 
\lambda_{\rm var}\sim 40~{\rm nm} \times
M_{6}^{0.32}\zeta^{0.45}\lambda^{-2.1}\alpha_{-1}^{-0.36}\beta_{-1}^{0.45}S^{-0.73}\theta_{0.2}^{-0.09}.
\eeq
Above, we have defined $\alpha_{-1}\equiv\alpha/0.1$,
$\beta_{-1}\equiv\beta/0.1$, and $\theta_{0.2}\equiv\theta/0.2$.
These equations already reveal that significant brightening will occur
primarily at photon energies in the hard UV to X-ray bands.  While the
characteristic frequency in eq.~(\ref{eq:hnuvar}) can move into the
optical bands for the most massive SMBHs, the overall evolutionary
timescale for these sources, eq.~(\ref{eq:tvisc}), becomes long.  The
most rapid evolution takes place as the cavity fills; once it is
filled, the system evolves more gradually to approach a standard
steady thin accretion disk solution around a single SMBH, with the
quantity $\nu L_{\nu}$ peaking at a frequency of roughly $\nu_{\rm
peak}\sim 15\nu_{\rm var}$.  The spectrum falls off steeply at higher
frequencies, and is likely unobservable above $\nu\gtrsim 3\nu_{\rm
peak}$.

Prior to decoupling, the luminosity of the disk at frequencies
$\nu\gtrsim \nu_{\rm var}$ is negligible.  Once the cavity is filled,
the monochromatic luminosity at frequencies $\nu_{\rm var}\ltsim
\nu\ltsim \nu_{\rm peak}$ reaches
\beq
\nu L_{\rm \nu}\sim 5-30\times 10^{42}\erg\,\s^{-1}
M_{6}^{0.92}\zeta^{-0.42}\lambda^{-1.7}\alpha_{-1}^{0.34}\beta_{-1}^{-0.42}S^{1.2}\theta_{0.2}^{0.08}.
\label{eq:nuLvar}
\eeq
In order to be identifiable in a survey, an afterglow source must
exhibit significant brightening, at least comparable to the typical
variability of typical AGN, at luminosities and frequencies to which
the survey is sensitive.

\subsection{Modeling the Population of Afterglow Sources}

We now turn to estimating the number of identifiable afterglow
sources.  The approach described below closely follows that described
in \cite{HKM09} for estimating the number of pre-merger sources that
may be detectable by their periodic variability.  We begin by
prescribing the quasar luminosity function as a proxy for the SMBH
mass function.  Specifically, we adopt the bolometric luminosity
function of \cite{Hopkins+07b}, and suppose that during a typical
bright quasar phase, the luminosity and SMBH mass are related via a
simple approximate relation, $L(M)\sim f_{\rm Edd}L_{\rm Edd}(M)$,
where $L_{\rm Edd}(M)$ is the Eddington luminosity for an object with
mass $M$, and $f_{\rm Edd}$ is a constant.  This is an admittedly
rough estimate, as $f_{\rm Edd}$ is known to have a non-negligible
spread among the population of bright quasars.  However, our simple
estimate is sufficient for a proof-of-concept; a more precise
calculation is not warranted, given the approximate and highly
idealized nature of the afterglow model and the uncertainty in the
system parameters.  We choose the fiducial values $f_{\rm Edd}\sim
S\sim 0.3$ \citep[e.g.,][]{Kollmeier+06} and assume a rest-frame
quasar lifetime of $t_{\rm Q}\sim 10^{7}\yr$
\citep[e.g.,][]{Martini04}.

We further assume that there is a one-to-one correspondence between a
SMBHB merger and a bright quasar episode, i.e. that a SMBHB merger
ultimately triggers quasar activity.  This assumption is consistent
with our post-merger disk evolution model, which naturally leads to a
state with a fully extended disk around a single SMBH, as long as fuel
remains available to maintain near-Eddington accretion rate at the
outer edge of the disk.  Given the comoving number density of AGN
$dn_{\rm AGN}/dM$, we are interested in the subset of SMBH merger
remnants that are producing an observable, brightening afterglow, and
have not yet settled to a later, presumably steadier quasar phase.  To
estimate this fraction of SMBHs, we use the afterglow model described
above, and calculate the duration $t_{\rm ag}$ over which the
photometric emission from a circumbinary disk brightens at a rate
exceeding some threshold value. This threshold should be chosen to
correspond to a brightening rate that is not only measurable, but is
also distinguishable from other possible sources of time-variability.
The number of variable sources $N_{\rm ag}$ in the sky of such SMBHB
remnants in a given mass and redshift range is then
\beq
N_{\rm ag}\sim 
\Delta V(z_{\rm lo}, z_{\rm hi})\int
\frac{dn_{\rm AGN}}{d\ln M}\left[\frac{t_{\rm ag}(M)}{t_{\rm Q}}\right]
~d\ln M,
\label{eq:nag}
\eeq
where $\Delta V(z_{\rm lo}, z_{\rm hi})$ is the cosmological comoving
volume between redshifts $z_{\rm lo}$ and $z_{\rm hi}$ and $n_{\rm
AGN}$ is the space density of SMBHs of mass $M$, inferred from the
quasar luminosity function evaluated at $L(M)$.  It is worth
emphasizing that this expression does not assume that the birthing
quasars have a bolometric luminosity of $L(M)$ -- rather, $L(M)$ here
represents the characteristic luminosity that is produced by SMBHs of
mass $M$ in their quasi--steady quasar state; this asymptotic
luminosity is reached only $\gsim 100$ yrs after the merger, according
to our afterglow models.

\section{Results and Discussion}

\subsection{Basic Parameter Dependencies}

The dependence of the number of detectable variable sources on the
various model parameters for the binary-plus-disk population is of
obvious interest, and is not trivial, as each parameter affects
differently the luminosity, spectral frequency and brightening rate of
the afterglow.  For example, increasing the binary mass $M$ increases
the source luminosity and lowers the characteristic frequency of the
source, while making the afterglow evolve more slowly --- thus, the
brightening and hardening rates both decrease (making identification
more difficult) while the total flux and the total afterglow lifetime
both increase (making a detection easier).

\begin{figure}
\centerline{\hbox{
\plotone{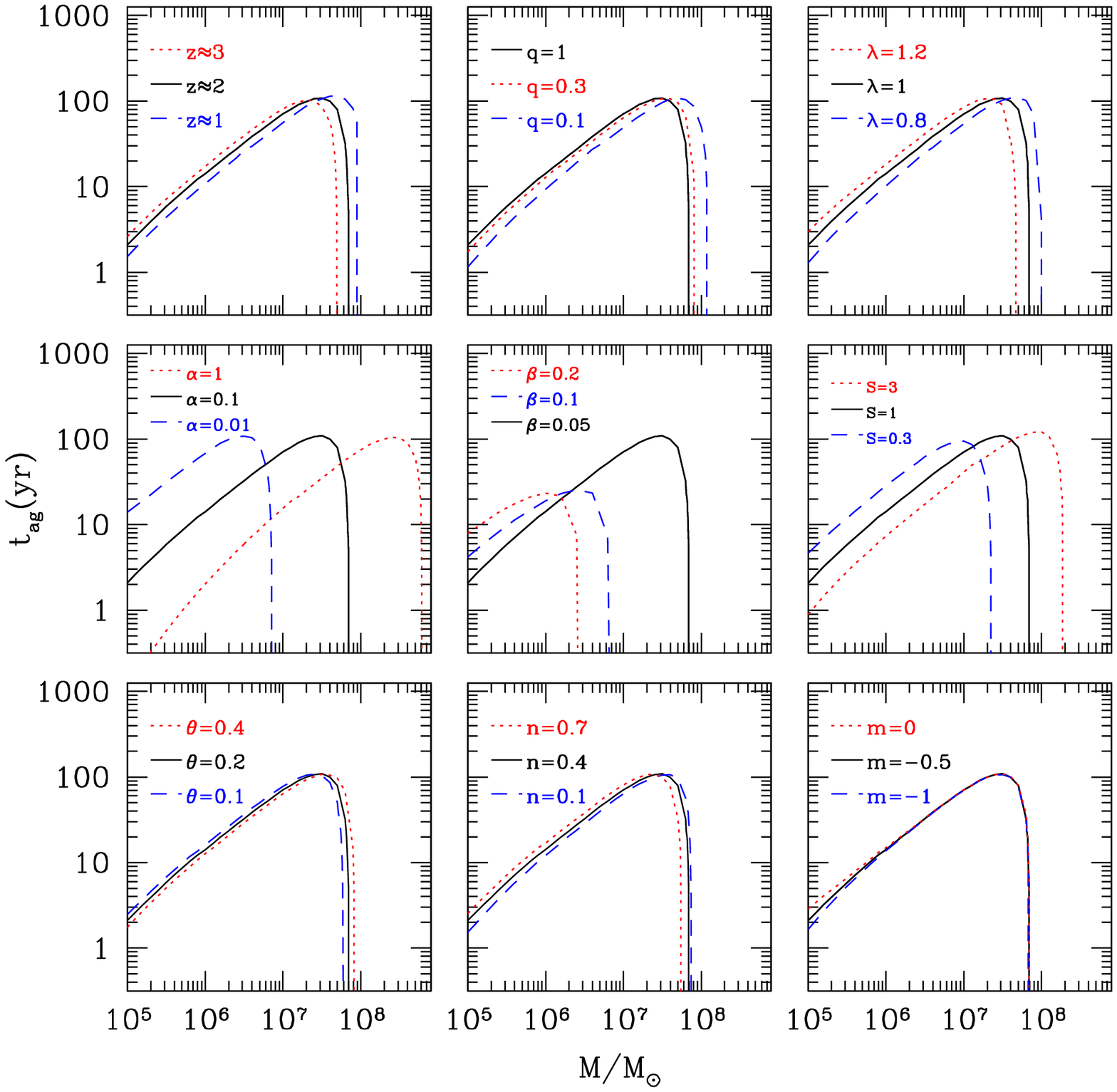}
}}
\caption{The approximate amount of time spent by an afterglow source
  with an observable band brightening rate of at least $10\%
  ~\yr^{-1}$ in the $0.1-3.5$ keV photon energy range.  The most
  luminous such sources spend $t_{\rm ag}\gtrsim 10\yr$ in this
  brightening phase, roughly independently of the model parameter
  values. The solid curve in each panel shows the fiducial set of
  parameters, $q=\alpha_{-1}=\lambda=S=\theta_{0.2}=1$, $\beta=0.05$,
  $z=2$, $n=0.4$ and $m=-1/2$.  The dotted and dashed curves show how
  the duration of the rapidly brightening phase is affected by changes
  in each of the model parameters (see \S~2.1 for a definition of each
  parameter).  }
\label{fig:tagm}
\end{figure}

To illustrate how each of the parameters and the source redshift
affect the detectability of variable afterglow sources, in
Figure~\ref{fig:tagm} we first plot the basic quantity $t_{\rm ag,
obs}=(1+z)t_{\rm ag}$, representing the amount of time sources are
observed to spend at or above the required threshold for the
brightening rate.  The threshold in this figure is set at $d\ln L_{\rm
X}/dt_{\rm obs}$ of at least $10\% \yr^{-1}$, in the soft X--ray
frequency window of $0.1-3.5$ keV (motivated by the proposed all-sky
X-ray transience survey {\it LWFT}; see below).  The solid black curve
in each panel shows $t_{\rm ag, obs}$ for the fiducial parameter
values $q=\alpha_{-1}=\theta_{0.2}=S=\lambda=1$, $\beta=0.05$,
$n=0.4$, $m=-1/2$, and $z=2$.  In each panel, we show how the apparent
duration of the rapidly brightening phase is affected by changes
(dashed and dotted lines) in one of the system parameters.

\begin{figure}
\centerline{\hbox{
\plotone{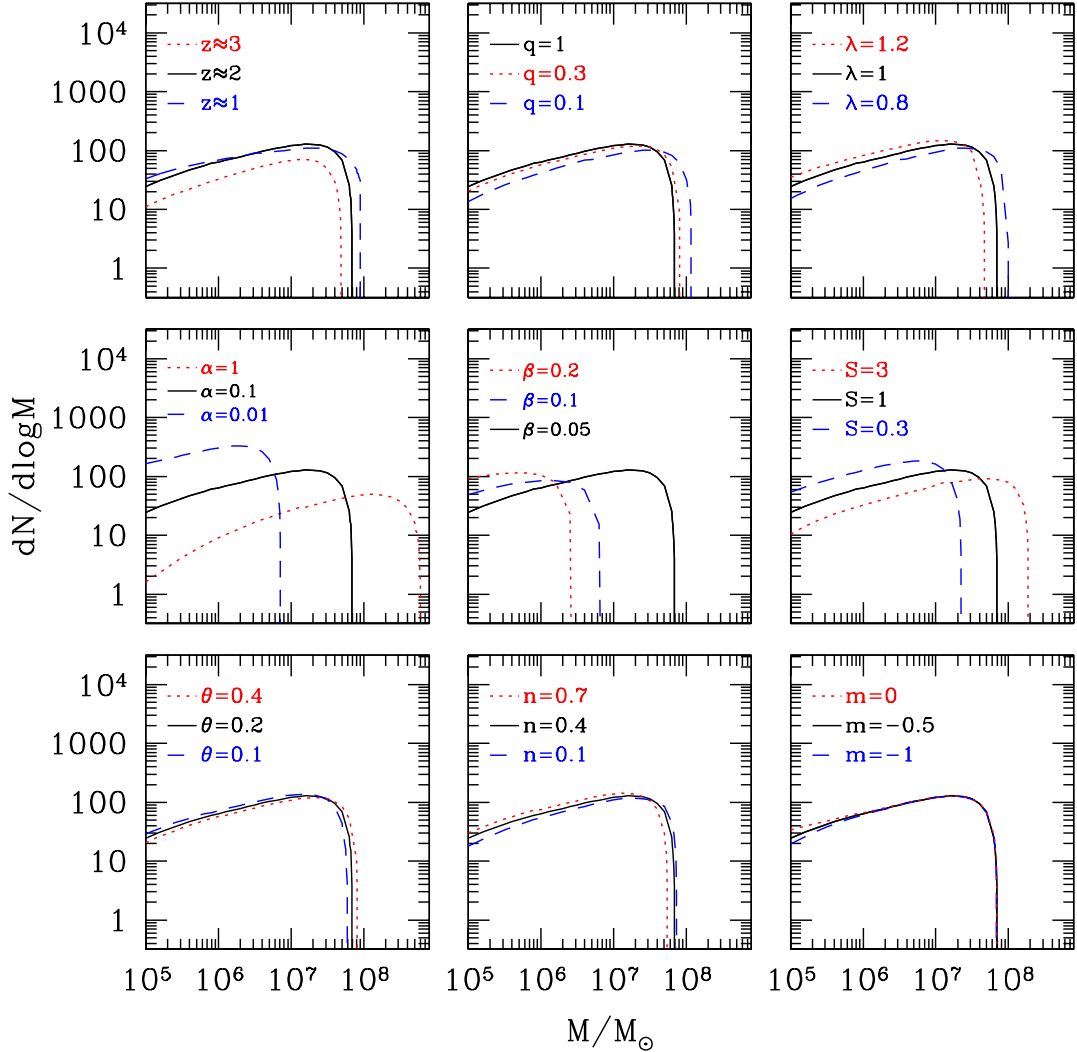}
}}
\caption{The approximate number of afterglow sources in the sky at any
  given moment whose luminosity in the $0.1-3.5$ keV energy range is
  (1) at least $10^{40}\erg\,\s^{-1}$ and (2) increasing by at least
  $10\% ~\yr^{-1}$.  We use the AGN luminosity function of
  \cite{Hopkins+07b} as a proxy for the SMBH mass function, and
  associate each episodic activity of AGN with a SMBH merger whose
  afterglow light curve is given by the time-dependent model of
  \cite{TM10}.  The solid curve in each panel shows the fiducial set
  of parameters, $q=\alpha_{-1}=\lambda=S=\theta_{0.2}=1$,
  $\beta=0.05$, $1.5<z< 2.5$, $n=0.4$ and $m=-1/2$.  The dotted and
  dashed curves show how the number of sources are affected by changes
  in each of the parameters.  Note that the most massive sources are
  not the most luminous in the band of interest (see text).  For
  plausible parameter values, the most luminous sources have masses,
  band luminosities, and number in the sky of $\gtrsim 10^{6}\Msol$,
  $L\gtrsim 10^{43}\erg\,\s^{-1}$, and overall number $N_{\rm
    ag}\ltsim 100$, respectively.  The masses of luminous sources of
  interest coincide with the {\it LISA}
  sensitivity window, $10^{5}-10^{7} (1+z)^{-1}\Msol$.  }
\label{fig:dndm}
\end{figure}

In Figure \ref{fig:dndm}, we plot the corresponding number $dN_{\rm
ag}/d\ln M$ of sources that exhibit a band luminosity $L_{\rm X}$ of
at least $10^{40}\erg\,\s^{-1}$ and an observed brightening rate $d\ln
L_{\rm X}/dt_{\rm obs}$ of at least $10\% \yr^{-1}$ in the same
$0.1-3.5$ keV frequency window.  This is given by the product of the
duty cycle $t_{\rm ag, obs} / t_Q$ and the space density of SMBHs
(eq.~\ref{eq:nag}), except that a further cut is imposed in
luminosity.  This is because the brightening rate initially may exceed
the threshold for SMBHs with masses near the low--mass end of the range
shown in the figure, but their band fluxes are still below the imposed
threshold; these sources are excluded by subtracting the duration of
this initial, sub-luminous state from the duty cycle $t_{\rm ag, obs}
/ t_Q$.  The number is computed for the whole sky, and over a redshift
range $1<z<3$.  The line-style scheme is the same as in Figure
\ref{fig:tagm}: the solid black curves show $dN_{\rm ag}/d\ln M$ for
fiducial parameter values, and the dashed and dotted curves show how
the number of rapidly brightening sources depends on each
parameter. The optimistic luminosity and brightening thresholds in
this figure are chosen purely for demonstrative purposes.  For
reference, a source with $L_{\rm X}\sim 10^{40}\erg\,\s^{-1}$ and
$z\sim 1$ would, in fact, have a flux of only $\ltsim
10^{-18}\erg\,\s^{-1}\cm^{-2}$, and thus be too faint to be monitored
for variability.  Also, AGN have been observed to vary in their X-ray
brightness by as much as order unity on timescales of years.  Although
the afterglows in question would exhibit a monotonic increase in X-ray
brightness, along with a corresponding monotonic spectral hardening, it
is unclear whether a brightening rate of $10\% \yr^{-1}$, even if
sustained for several years, and accompanied by a monotonic hardening
of the spectrum, would be sufficient to distinguish an afterglow
candidate from other X-ray variable sources.

The steep cutoff at high binary masses seen in both $t_{\rm ag}$ and
$dN_{\rm ag}/d\ln M$, in Figures \ref{fig:tagm} and \ref{fig:dndm},
respectively, has two causes.  One is that for sufficiently large $M$,
the emission frequency of the source becomes too low, and falls out of
the soft X--ray window.  The other reason is that the disk evolution
timescale $t_{\rm visc}$ scales as $\propto M^{1.3}$, so that for
sufficiently large $M$ the disk evolves so slowly that its brightening
rate never reaches the adopted threshold of $10\% \yr^{-1}$.

Figures \ref{fig:tagm} and \ref{fig:dndm} also show that the duration
of the brightening phase, and the mass function of the afterglow
sources depend strongly only on the parameters $\alpha$, $\beta$ and
$S$.  This is due to the fact that the afterglow frequency range and
evolution timescale scale steeply with these parameters (see equations
\ref{eq:tvisc} and \ref{eq:hnuvar}).  Increasing $\alpha$, increasing
$S$ and decreasing $\beta$ relative to their fiducial values all have
the effect of increasing the afterglow emission frequencies and
pushing it further into the survey frequency window, while also
increasing the brightening rate of the afterglow.  The quantities
$\nu_{\rm var}$ and $t_{\rm visc}$ both depend only weakly on
$\theta$, and $\zeta$ varies too weakly in the range $0.1\ltsim q\le
1$ to have a sizable effect.  Increasing (decreasing) the parameter
$\lambda$ results in afterglows that are further inside (outside) the
frequency window but evolve much more slowly (quickly) -- the two
effects tend to cancel out, and yield a relatively weak overall effect
on $dN_{\rm ag}/d\ln M$.

It is worth noting that the brightest afterglow sources satisfying a
fixed $d\ln L_{\rm X}/dt_{\rm obs}>10\% \yr^{-1}$ are not the most
massive ones. This is because disks around more massive BHs evolve
more slowly, and the brightening rate is greatest early in the
post-decoupling disk evolution when the source is dimmer.  For most of
the parameter value combinations probed in Figure \ref{fig:dndm} ---
excepting $\alpha$, $\beta$ and $S$ for the moment --- the most
luminous sources brightening at or above the threshold rate are those
with binary masses of $(0.5-2)\times 10^{6}\Msol$.  Interestingly,
this mass range lies in the middle of {\it LISA}'s sensitivity window.

The maximum band luminosities of these sources are typically
$1-4\times 10^{43}\erg\,\s^{-1}$, and behave roughly as described in
equation \eqref{eq:nuLvar}.  For $\alpha=1$ and $S=3$, the masses and
luminosities for the brightest afterglow sources are somewhat greater:
$\gtrsim 5\times 10^{6}\Msol$ and $\gtrsim 10^{44}\erg\,\s^{-1}$.

The range of luminosities across the parameter combinations probed in
Figures \ref{fig:tagm} and \ref{fig:dndm} correspond to an observed
soft X-ray flux of $F_{\rm X} \sim10^{-16}-10^{-14} \times
S^{1.2}~{\rm erg~s^{-1}~cm^{-2}}$ in the range $1\ltsim z \ltsim 3$.
Thus, if the approximate location of the source is known via a prior
GW detection, the afterglow would be observable during the rapidly
brightening phase at the sensitivity achieved by existing instruments
such as {\it XMM-Newton}, {\it ROSAT HRI}, and {\it Chandra} \citep[at
$\sim 100$ ks exposure; see, e.g.,][]{Brandt05}.  For the parameter
combinations probed in Figures \ref{fig:tagm} and \ref{fig:dndm},
there are at most $\sim 100$ sources in the sky with $L_{\rm X}\gtrsim
10^{43}\erg\,\s^{-1}$ and $d\ln L_{\rm X}/dt_{\rm obs}>10\% \yr^{-1}$,
and they maintain this luminosity and brightening rate for $t_{\rm ag,
obs}\gtrsim 10\yr $ in the observer frame.

\subsection{Counts of Birthing Quasars in X-ray and Optical Surveys}

Applying the methods described above to estimate the afterglow
light-curve and source population, we next calculate the source counts
of identifiable afterglows as a function of their apparent flux.
Based on the fact that the mergers of massive dark matter halos peak
at a redshift $z\sim 2$, we limit our analysis to AGN in the redshift
range $1< z < 3$.  We assume that the typical disk-plus-binary system
has parameter values $q=0.1$, $\beta=0.05$, $\lambda=1$, $\theta=0.2$
$n=0.4$, and $m=-1/2$.  In reality, these parameters will vary from
system to system, perhaps by a great deal.  However, because our
calculations above (equation \ref{eq:nuLvar} and Figure
\ref{fig:dndm}) suggest that neither the luminosity nor the mass
function of afterglow sources are likely to be affected by the values
of these parameters by more than an order of magnitude, we limit
further exploration of the parameter space to the $\alpha$-$S$ plane.
The value of the viscosity parameter $\alpha$ is highly uncertain;
however, numerical simulations of MHD disks suggest in the
radiation-dominated regions of an accretion disk $\alpha\sim 1$ is
consistent with our type of viscosity prescription $\nu\propto P_{\rm
gas}$ \citep{HBK09}.  Assuming that the mass supply rate around the
afterglow phase is at least comparable to the rate during prolonged
AGN activity, we expect $S$ to be at least as great as the ratio of
the typical luminosity of the typical AGN episode to the Eddington
luminosity limit of the SMBH engine, i.e. $S\ge f_{\rm Edd}=0.3$.
Depending on how much mass the circumbinary disk accumulates near its
inner wall prior to decoupling, $S$ may, however, significantly exceed
$f_{\rm Edd}$.

\subsubsection{X-ray Surveys}

We consider a hypothetical survey with observational capabilities
similar to those of the proposed {\it LWFT} mission, sensitive to
photon energies of $0.1-3.5$ keV, down to fluxes of $\sim 3.5 \times
10^{-14}~{\rm \erg\,\s^{-1}~\cm^{-2}}$ (confusion limit for {\it
LWFT}).

In Figure~\ref{fig:fluxX}, we present the X-ray source counts of
objects increasing in band luminosity by at least $3 \%\yr^{-1}$, $10
\%\yr^{-1}$, $30\%\yr^{-1}$ and $100\%\yr^{-1}$ in the relevant
frequency band.  (Note that the energy range approximately coincides
with the sensitivity window of existing deeper-exposure telescopes
such as {\it Chandra} and {\it XMM-Newton}, and the planned all-sky
survey {\it eROSITA}.)  In the figure, the height of the histogram
pillars are the number of sources in each logarithmic flux bin, with
each bin having a width $\Delta \log_{10}(F\erg^{-1}\cm^{2}\s)=0.5$.
The dashed vertical line in each panel shows the approximate source
confusion flux limit.  These results suggest that only if $S$ is
large, i.e. if there is significant mass accumulation at the inner
wall of the circumbinary disk, then rapidly brightening afterglow
sources could be detected by the model mission: several sources at
$S\sim 3$ and as many as dozens of sources for $S\sim 10$.  The
brightest of these sources have central SMBH masses of $M\gtrsim
10^{6}S^{1.2}\Msol $, and thus most of the X-ray-detectable sources
would also be observable by {\it LISA.}  However, for $S\gtrsim 3$,
some would fall outside {\it LISA}'s sensitivity window.  These
sources are expected to continue to brighten at a slightly reduced
rate at harder frequencies, $h\nu> 3.5$ keV; this is a prediction that
could be tested with pointed follow-up observations.  Our calculations
indicate that birthing quasars will be difficult to identify with
existing and planned wide-angle soft-X-ray surveys.  For example, the
{\it eROSITA} all-sky survey is expected to have a semi-annual flux
limit of $\gtrsim 10^{-13}\erg\s^{-1}cm^{-2}$ with a time resolution
of $\sim$ months at those flux levels.  Intrinsic absorption of soft
X-rays by the birthing quasar itself could also be an observational
barrier for at least some of the sources \citep[see,
e.g.,][]{Brandt+00}, especially if galaxies harboring merging SMBHBs
tend to be more heavily shrouded in gas and dust than the general
population of active galaxies.

Whether the monotonic brightening of the afterglow would be
sufficiently distinguishable from the X-ray variability of faint AGN
and other sources is an open (and more difficult) question.  Luminous
X-ray AGN have been known to vary by as much as $\sim 10-100\%
\yr^{-1}$ \citep[e.g.,][]{Mushot+93}.  Any monotonic brightening in
the X-ray must be distinguished from other sources of intrinsic
variability, in addition to any instrumental error close to the
faint-flux detection limit.  It would appear likely, however, that a
source that monotonically increases its X-ray luminosity by up to
order unity per year for several years, while showing a consistent and
monotonic hardening of its spectrum for the entire duration, would be
fairly unusual.  While an increase in the accretion rate through a
standard thin accretion disk is also expected to produce a
simultaneous brightening and spectral hardening, the spectral
evolution will be different in detail.  We also note that a
variability survey, utilizing the ROSAT all sky survey, as well as
pointed ROSAT observations \citep{GTB01} found no correlation between
changes in flux and spectral hardening in a sample of luminous
soft-X-ray AGN.  If the brightening is caused by the viscous
afterglow, it would also slow down on humanly tractable timescales and
the light curve could be checked at different observational
frequencies against the evolutionary models for the viscously
spreading disk.  This feature would also help distinguish birthing
quasars from tidal disruption events.

It is worth emphasizing that a major caveat of the above analysis is
that the thin-disk formalism adopted in our afterglow model breaks
down for models with $S\gtrsim 3$.  Indeed, in this massive regime,
the disk midplane temperature becomes sufficiently high that a
one-dimensional estimate for the disk scale-height-to-radius ratio
$H/R$, evaluated at the cavity wall, exceeds unity \citepalias{MP05,
TM10}.  This suggests that horizontal advection would become a
significant factor in determining the structure of the inner accretion
flow.  The evolution and emission properties of the disk in this
advective regime is highly uncertain and it remains a subject of
active research.  Advective disks have generally been associated with
radiatively inefficient accretion states \citep[see][for a
review]{NQ05}, but the situation may be different here, with a high
density accretion flow and an intrinsically time-dependent flow.  As
suggested by \citetalias{MP05}, horizontal advection could act to make
the disk ``slim'', as in the models of \cite{Abramowicz+88}, and thus
relatively radiatively efficient.  It is also unclear how well binary
torques can open a central cavity in the first place, if the disk
becomes geometrically thick.  Adding to the various modeling
uncertainties is the possibility that a sizable fraction of the
viscously dissipated energy could be deposited into a hot disk corona
\citep[e.g.,][]{LMO03}, which would lead to a significant increase of
the afterglow high energy emission, well above what is predicted by
our strictly thermal afterglow models.

\subsubsection{Optical Surveys ({\it LSST})}

Our simple viscous emission model predicts that the same SMBH binary
remnant that produces an X-ray afterglow would brighten at lower,
optical frequencies at a rate of $\gtrsim {\rm a~few} \% \yr^{-1}$,
several years prior to the X-ray afterglow \citepalias{TM10}.  This
less pronounced variability is comparable in magnitude to the
intrinsic r.m.s. long-term variability of optical AGN (e.g.,
\citealt{Ulrich97}).  However, if the approximate location of the
source is known through a GW detection, searching for AGN exhibiting
steady optical/ultraviolet brightening may identify the source before
the X-ray afterglow.  Alternatively, in the absence of a GW signal, a
wide-angle variability survey of optical AGN with high time
resolution, such as those possible with {\it LSST}, could possibly
still be used to select afterglow candidates for follow-up X-ray
observations.

Motivated by these possibilities, we apply our simple source-count
estimate to the {\it LSST} {\it u} photometric band ($330 - 400$ nm).
We choose this band because it is the {\it LSST} filter with the
highest frequency range, and thus the one in which the afterglow
brightening is likely to be the most prominent.  The results are
presented in Figure \ref{fig:fluxu}.  In both panels, the dashed
vertical lines demarcate the limiting flux for signal-to-noise of $50$
over a $\sim 1\yr$ period ($\sim 450\s$ accumulated exposure).  With
the fiducial parameter choices in the left panel, there is clearly no
hope of a detection.  We also find that, unlike in the X-ray bands,
raising the value of $S$ does not increase the source counts.  This is
because while increasing $S$ beyond the fiducial value pushes the
evolving portion of the spectrum into the model instrument band, it
pushes it further out of the {\it LSST u} band, reducing the optical
variability.  Therefore, we instead show, in the right panel, the
expected number counts for the larger value $\alpha=1$, on which the
emission frequency depends less strongly.  We find that the sources
are still likely to be too dim and too few to be identifiable with
high confidence from among the large number of AGN expected to be
detected by {\it LSST}.  However, gradually brightening optical AGN
could still be cataloged, and their X-ray luminosities could be
cross-checked with data from instruments such as {\it eROSITA} for
subsequent X-ray afterglows, which may still prove useful.

More promising for optical surveys is the possibility that the X-ray
afterglow may be promptly reprocessed into optical or infrared
frequencies, either by dust surrounding the source \citepalias{MP05}
or by warps and vertical splaying in the circumbinary disk
\citepalias{TM10}.  If a significant portion of the X-ray emission is
reprocessed, then the afterglow source will appear as an AGN whose
optical/infrared luminosities can brighten by $d\ln
{L}_{\nu}/dt\gtrsim 10\%~\yr^{-1}$.  Such a monotonic variability
would exceed the typical long-term r.m.s. variability in AGN
brightness at these frequencies, and is likely to be identifiable by
{\it LSST}.  For purely demonstrative purposes, we adopt here a simple
toy model, in which a fraction $f_{\rm rp}$ of the emitted power above
$\nu>1$ keV in the source's rest frame is reprocessed (thermalized)
and re--emitted at frequencies below the ultraviolet energy $10$ eV,
so that the power below $h\nu \le 10$ eV is enhanced as
\beq 
L^{(\rm rep)}_{h\nu < 0.01 {\rm ~keV}}\sim L_{h\nu < 0.01 {\rm
    ~keV}}+f_{\rm rp} L_{h\nu > 1 {\rm ~keV}}.  
\eeq
We further assume that the fractional energy enhancement is roughly
uniform in the optical and infrared --- i.e., we ignore line emission
from recombination processes --- so that for our purposes the optical
spectral emission is given by
\beq 
\nu L_{\nu}^{(\rm rep)}\sim \nu L_{\nu} \times
\left(1+f_{\rm rp}\frac{L_{h\nu > 1 {\rm ~keV}}}{L_{h\nu < 0.01 {\rm
        ~keV}}}\right).  
\eeq
This prescription is similar in spirit to the model of
\cite{Gierli+09}, who showed that reprocessing signature of the X-ray
outbursts of the stellar-mass black hole system XTE J1817-330 is
consistent with reprocessing a constant fraction of the bolometric
X-ray luminosity.  We also neglect the reprocessing time, which is
dominated by the light-travel time and much shorter than the
variability timescales of interest here \citep[see,
e.g.,][]{Peterson2004}.

The source counts in the {\it LSST u} band for this model, for a
reprocessing fraction $f_{\rm rp}=10^{-2}$, are shown in Figure
\ref{fig:fluxrep}.  Our simple calculations suggest that perhaps
dozens of afterglows could be detected if the X-ray emission is
reprocessed, for moderately optimistic parameter values, e.g., for
$S\gtrsim 1$.  It is worth cautioning that the reprocessed fraction
$f_{\rm rp}$ is highly dependent on the vertical disk geometry, which
itself may be rapidly evolving during the afterglow.  For instance,
\citetalias{TM10} found in their non-irradiated afterglow models that
the scale-height-to-radius ratio $H/R$ can be a steeply increasing
function of radius during the period when the cavity is refilling, but
not necessarily before or after this phase.  Such complications should
be included, along with details of radiative transfer, in a more
realistic analysis of disk irradiation and reprocessing.  Finally,
absorption of the reprocessed UV/optical emission by gas and dust,
surrounding the nuclear SMBH on larger scales, could be another
observational hindrance for at least some of the sources.

\section{Conclusions}

Using an idealized model for the population of coalescing SMBHBs, and
for the light curve of the afterglow produced by the viscously
spreading post-merger circumbinary disk, we have shown that ongoing
afterglows of SMBHB mergers may be present in the data sets of wide
X-ray and optical surveys.  In soft X-ray bands, this requires that
the surface density and the viscosity in the circumbinary disk be at
the high end of the expected range, while afterglows could only be
found in optical surveys if the X-ray emission is promptly and
significantly reprocessed into optical frequencies.

\begin{figure}
\centerline{\hbox{
\plotone{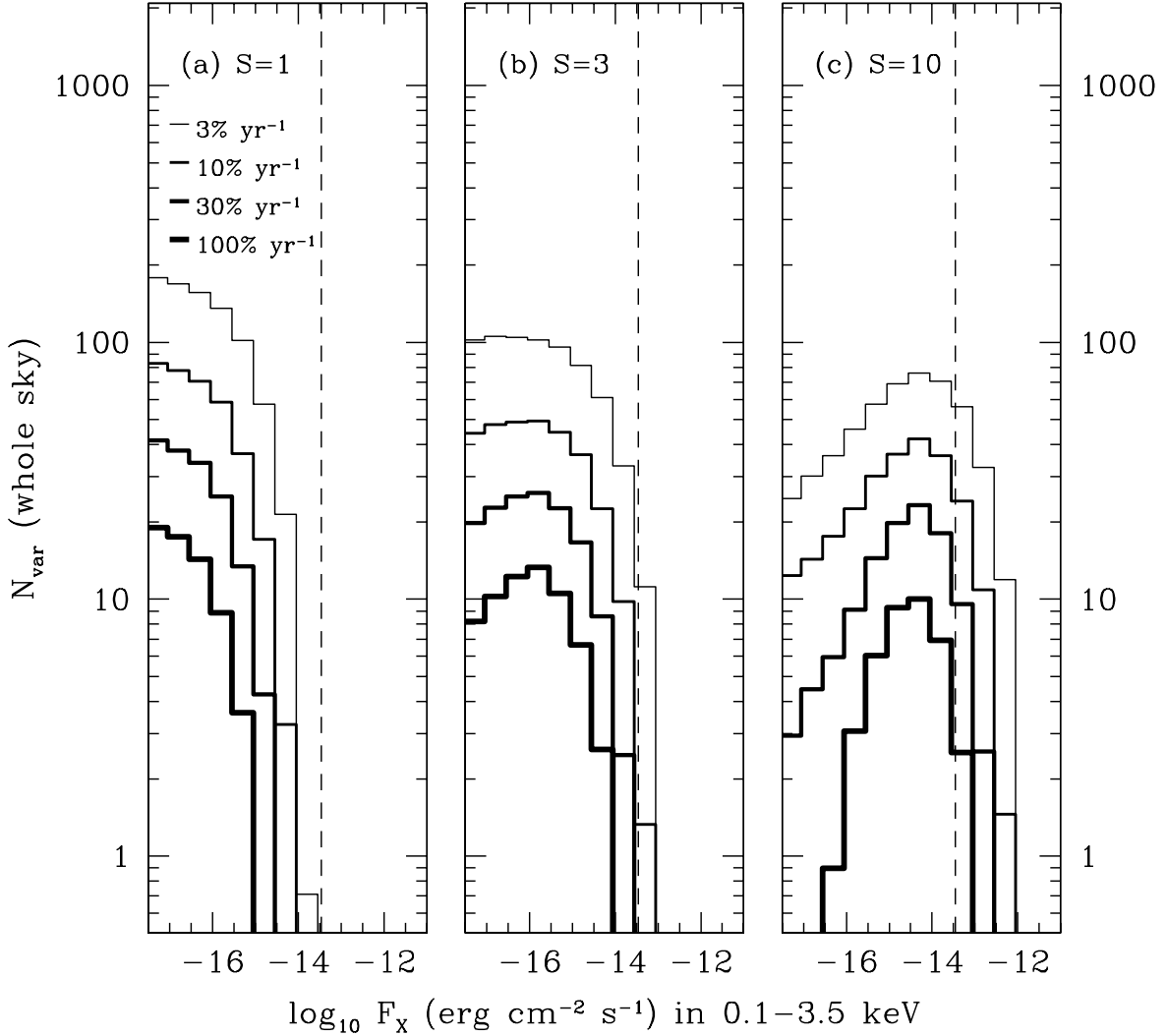}
}}
\caption{The number of afterglow sources as a function of their
  $0.1-3.5$ keV flux, in the redshift range $1<z< 3$.  The histogram
  shows the number of sources in each logarithmic flux bin of width
  $\Delta \log_{10}(F\erg^{-1}\cm^{2}\s)=0.5$.  We consider sources
  with $10^{6}\Msol< M< 10^{9}\Msol$.  The histograms demarcate, in
  order of decreasing line thickness, the number of sources in each
  flux bin exhibiting a brightening rate of at least $3\%~\yr^{-1}$,
  $10\%~\yr^{-1}$, $30\%~\yr^{-1}$, and $100\%~\yr^{-1}$.  The dashed
  vertical line in each panel is the detection limit for point sources
  in our model survey.  All panels have parameter
  values $q=0.1$, $\alpha_{-1}=\lambda=\theta_{0.2}=1$, $\beta=0.05$,
  $1.5<z< 2.5$, $n=0.4$ and $m=-1/2$.  We vary the value of $S$, a
  measure of the disk mass, in each panel: $S=1$ in panel (a), $S=3$
  in panel (b), and $S=10$ in panel (c).  }
\label{fig:fluxX}
\end{figure}

\begin{figure}
\centerline{\hbox{
\plotone{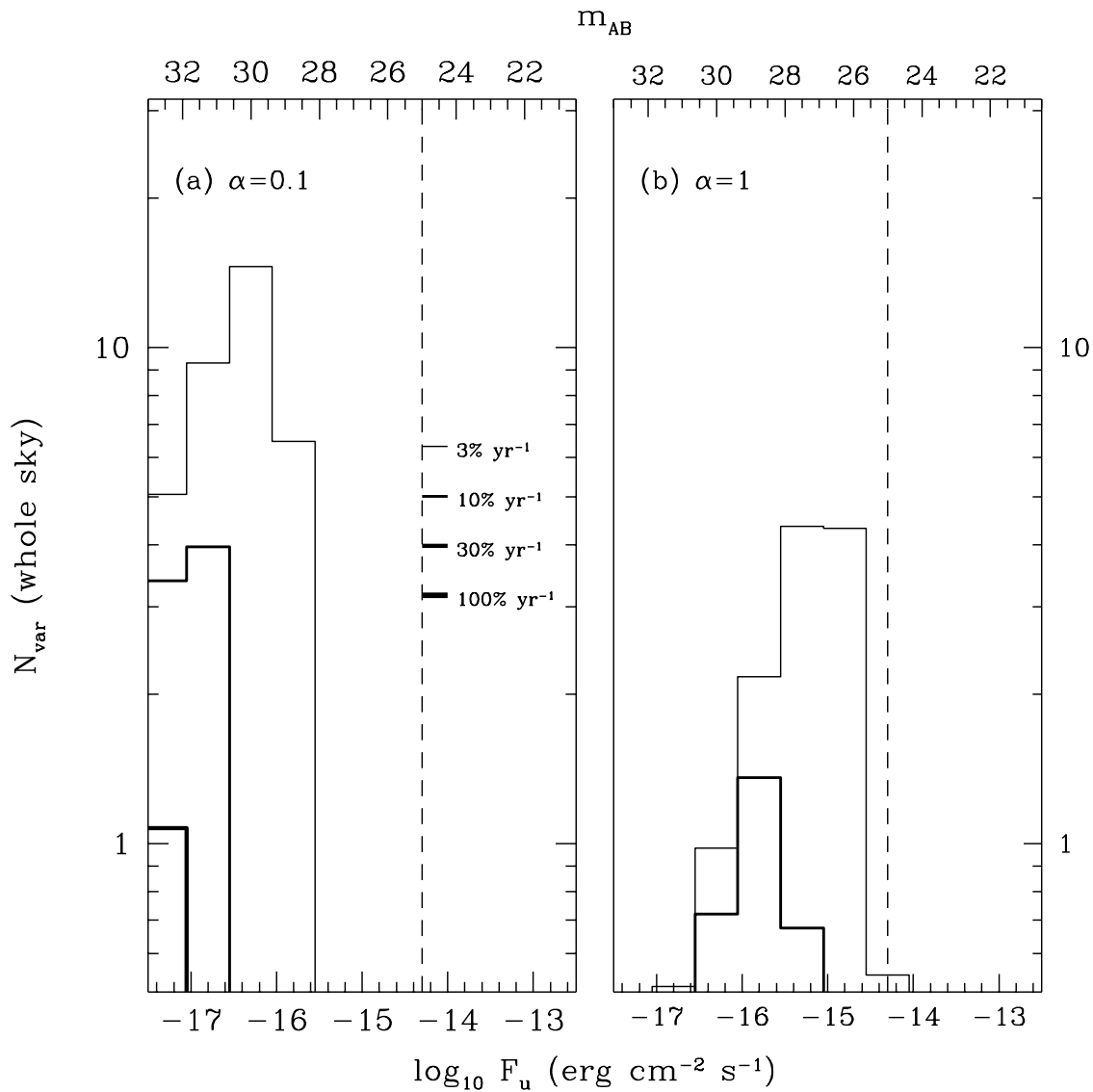}
}}
\caption{Same as Figure \ref{fig:fluxX} but in the {\it u} photometric
band ($330-400$ nm).  The dashed vertical line is the limiting flux to
achieve a signal-to-noise of at least $50$ over $\sim 1\yr$ of
operation for {\it LSST} ($\sim 30$ exposures at $15\s$ each).  Both
panels have parameter values $q=0.1$, $S=\lambda=\theta_{0.2}=1$,
$\beta=0.05$, $1<z< 3$, $n=0.4$ and $m=-1/2$.  Panel (a) has
$\alpha=0.1$, and panel (b) has $\alpha=1$ (more viscous disk).  }
\label{fig:fluxu}
\end{figure}

\begin{figure}
\centerline{\hbox{
\plotone{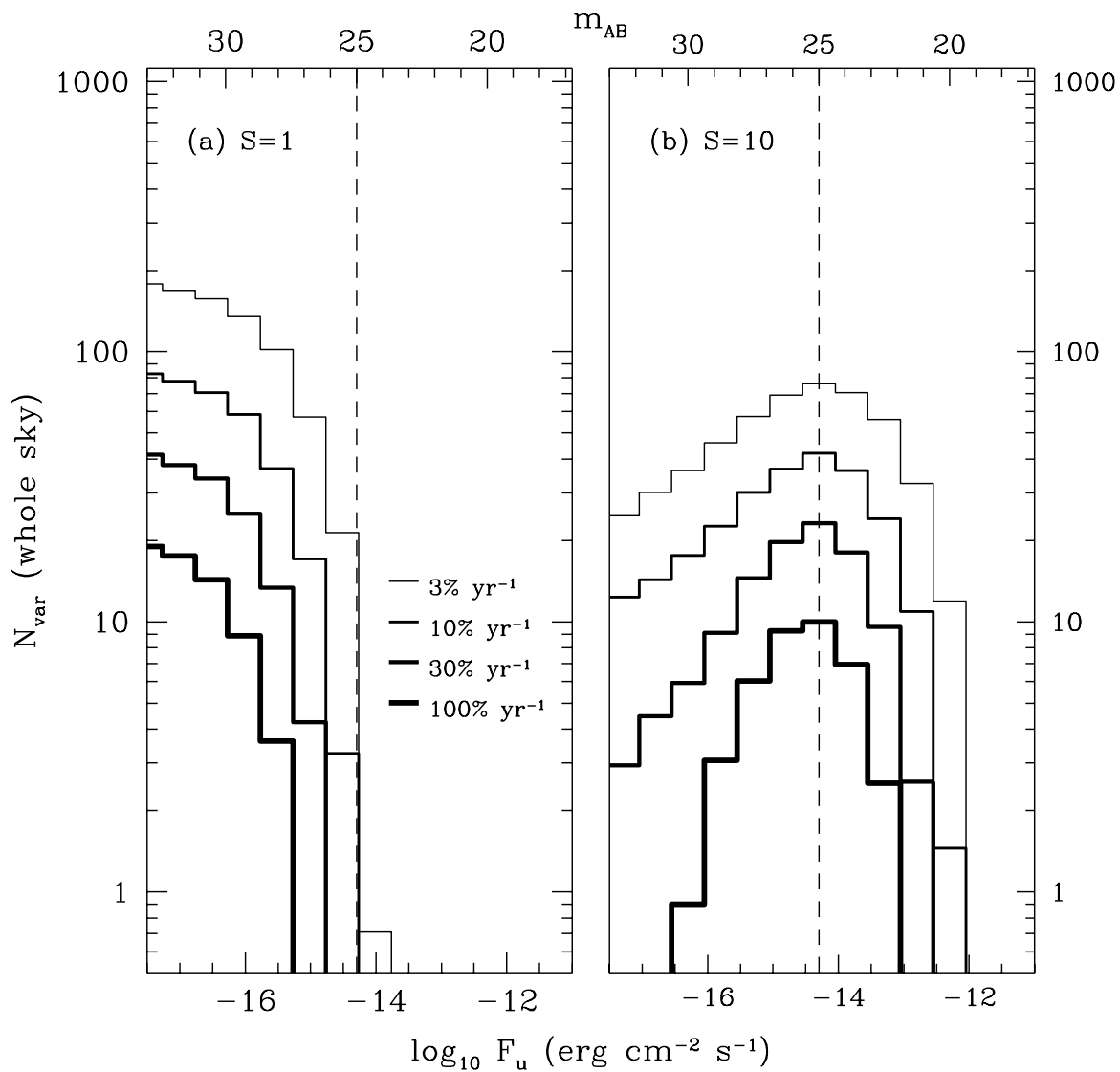}
}}
\caption{Same as Figure \ref{fig:fluxu} but assuming that $f_{\rm rp}=1\%$
  of the power emitted above 1 keV is reprocessed into optical and
  infrared bands (see text for details).  The dashed vertical line is
  the limiting flux as in Figure~\ref{fig:fluxu}.  Both panels have
  parameter values $q=0.1$, $\alpha_{-1}=\lambda=\theta_{0.2}=1$,
  $\beta=0.05$, $1< z<3$, $n=0.4$ and $m=-1/2$.  Panel (a) has $S=1$,
  and panel (b) has $S=3$ (more massive disk).  }
\label{fig:fluxrep}
\end{figure}

Despite the highly approximate nature of our analysis and other model
uncertainties, our calculations provide a proof-of-concept for a very
general hypothesis: {\it SMBHB mergers may exhibit identifiable,
steady brightening rate for a period of the order of decades, and such
afterglows could be detected serendipitously in a large survey that
revisits the sky at least every few months for several years.}  Our
more specific findings can be summarized as follows:
 
\begin{itemize}

\item For optimistic parameter values, several birthing quasars,
brightening by at least $d\ln L_{\rm X}/dt_{\rm obs}>30\% \yr^{-1}$
for several years, could be identified in the 0.1 - 3.5 keV soft X-ray
band by an all-sky survey with specifications comparable to those
proposed for the {\it LWFT} mission.

\item At any given time, there could be up to $N_{\rm ag}\sim 100$
sources in the sky that exhibit a brightening at or above $d\ln L_{\rm
X}/dt_{\rm obs}>10\% \yr^{-1}$, with soft X-ray luminosities $L_{\rm
X}\gtrsim 10^{42}\erg\,\s^{-1}$.  The most luminous sources typically
spend $t_{\rm ag, obs}\gtrsim 10\yr$ in this state, and thus can be
monitored on humanly tractable timescales.  These numbers depend
weakly on most system parameters.

\item To have any hope of detecting birthing quasars, a survey has to
reach a depth of at least a few $\times
10^{-13}\erg\,\s^{-1}\cm^{-2}$.  However, the slopes of our calculated
$\log N - \log S$ distributions at fluxes just below this threshold
are relatively shallow (Figures \ref{fig:fluxX}, \ref{fig:fluxu} and
\ref{fig:fluxrep}), implying that surveys should favor large angular
sky coverage over depth, once they reach this flux threshold.

\item If identified, candidate sources can be followed up by pointed
observations at higher frequencies, where they are expected to
continue both their monotonic brightening and their spectral
hardening.

\item Most birthing quasars that are identifiable have,
coincidentally, SMBH masses lying in the middle of {\it LISA}'s
sensitivity window ($M\sim 10^{6}\Msol$), and are thus members of the
same population that would be probed with GW detections.
However, a minority ($\gtrsim {\rm few}~\%$ for $S\gtrsim 3$)
of the detectable X-ray variables have masses
of $\gsim 10^7~{\rm M_\odot}$, probing a population above {\it LISA}'s
range.

\item These sources may be identifiable by {\it LSST} if a fraction as
low as $\sim1\%$ of the X-ray flux is promptly reprocessed into the
optical frequencies.

\end{itemize}

Our calculations are contingent on theoretical caveats of the
afterglow scenario we have considered.  The two primary uncertainties
regarding the post-merger evolution of the circumbinary cavity are
related to the viscous and advective properties of the disk.  As
stated in \S~\ref{subsec:curve}, the viscosity of accretion flows,
including the possibility of viscous instability, are not well
understood when radiation pressure dominates gas pressure, which is
the relevant regime for the gas refilling the circumbinary cavity.
Additionally, the disk may be geometrically thick \citepalias{MP05,
TM10}, either right at decoupling or later during the afterglow phase,
suggesting that horizontal advection may play a significant role in
determining the surface density evolution and the disk net emission
properties.  The importance of viscous instabilities in
radiation-dominated accretion flows remains a general open question,
and the role of advection in a viscously spreading accretion flow
remains a largely unexplored regime.  More detailed studies of the
circumbinary cavity will be needed to address how these effects may
affect the emission predicted by simple analyses based on a thin disk
formalism such as ours.  Another major uncertainty is the validity of
our assumption that quasar activity can be associated with SMBH
coalescence.  In reality, there may not be a one-to-one relation: it
is possible that for at least some AGN, gas accretion or changes in
radiative efficiency are triggered by mechanisms other than SMBH
mergers; conversely, some SMBH mergers may not trigger prolonged
quasar activity.  If the former is true, our analysis overestimates
the number of identifiable afterglow sources; if the latter is true,
then our results could in principle be an underestimate.

For completeness, we note that while we focused here on the viscous
afterglows, other SMBHB merger--related signatures could also be
looked for in EM surveys.  For example, the GW--emission--induced
mass--loss and recoil can cause strong disturbances in the
circumbinary disk, which can produce a detectable afterglow
\citep{Lippai+08, SK08, SB08, O'Neill+09, Megevand+09, Corrales+10,
Rossi+10}.  For the low SMBH masses of $\sim10^6\Msol$ relevant for {\it
LISA}, these signatures are expected to have a short duration $\sim$
few years (e.g., \citealt{Corrales+10}) and would be too rare to be
found serendipitously, without a trigger from {\it LISA}. However
\citet{SK08} and \citet{SB08} focused on these signatures in disks
around more massive SMBHs, which occur on longer ($\sim 10^4$yr)
time--scales, and proposed detecting a flare by monitoring a
population of AGN in the infrared or X-rays bands.  Another
possibility is that the binary is activated, and produces periodic
emission, tracking the orbital frequency, prior to the
merger. \citet{HKM09} argued that as long as this emission is at a few
percent of the Eddington luminosity, a population of these variable
sources, with periods of tens of weeks, may be identifiable in optical
or X-ray surveys.

To conclude, the concomitant observation of a SMBHB merger based
on GW and EM signals remains by far the most promising scenario for
the unambiguous detection of such systems. The precision with which
{\it LISA} would determine the masses, spins, and luminosity distances
of coalescing binaries can not be replicated by current or planned EM
telescopes.  However, detections based on EM signatures alone could
still help identify SMBHB mergers before {\it LISA} is launched, and
perhaps more importantly, possibly outside {\it LISA}'s mass
sensitivity window.  Detecting the EM signatures from the mergers of
the most massive SMBHs would complement the synergistic EM-plus-GW
observations of lower-mass systems, and help provide a more complete
picture of the accretion physics and cosmological evolution history of
SMBHBs.

$~$

TT thanks Joshua Peek and Jennifer Sokoloski, and ZH thanks Stefanie
Komossa, Jules Halpern, and Richard Mushotzky, for useful
conversations on AGN surveys.  This work was supported by the
Pol\'anyi Program of the Hungarian National Office for Research and
Technology (NKTH) and by NASA ATFP grant NNXO8AH35G.

\bibliographystyle{apj}
\bibliography{paper}

\end{document}